\documentclass[pdflatex,sn-nature]{sn-jnl}

\usepackage{graphicx}%
\usepackage{multirow}%
\usepackage{amsmath,amssymb,amsfonts}%
\usepackage{amsthm}%
\usepackage{mathrsfs}%
\usepackage[title]{appendix}%
\usepackage{xcolor}%
\usepackage{textcomp}%
\usepackage{manyfoot}%
\usepackage{booktabs}%
\usepackage{algorithm}%
\usepackage{algorithmicx}%
\usepackage{algpseudocode}%
\usepackage{listings}%

\usepackage{csquotes}
\MakeOuterQuote{"}

\usepackage{bm}
\usepackage{braket}
\renewcommand{\v}[1]{\bm{\mathrm{#1}}}
\newcommand{\m}[1]{\bm{\mathsf{#1}}}

\begin{document}

\title[Article Title]{Ultrafast all-optical generation of pure spin and valley currents}

\author*[1]{\fnm{Deepika} \sur{Gill}}\email{deepikagill1008dg@gmail.com}

\author*[1,2]{\fnm{Sangeeta} \sur{Sharma}}\email{geet1729@gmail.com}

\author*[1]{\fnm{Sam} \sur{Shallcross}}\email{phsss75@googlemail.com}

\affil[1]{Max-Born-Institute for Non-Linear optics, Max-Born Strasse 2A, 12489 Berlin, Germany}
\affil[2]{Institute for theoretical solid-state physics, Freie Universit\"at Berlin, Arnimallee 14, 14195 Berlin, Germany}

\abstract{Pure currents comprise the flow of a two state quantum freedom -- for example the electron spin -- in the absence of charge flow. Radically different from the charge currents that underpin present day electronics, in two dimensional materials possessing additional two state freedoms such as valley index they offer profound possibilities for miniaturization and energy efficiency in a next generation spin- and valleytronics. Here we demonstrate a robust multi-pump lightwave protocol capable of generating both pure spin and valley currents on femtosecond times. The generation time is determined by the 2d material gap, with the creation of pure spin current in WSe$_2$ at 40~fs and pure valley current in bilayer graphene at $\sim$200~fs. Our all-optical approach demands no special material design, requiring only a gapped valley active material, and is thus applicable to a wide range of 2d materials.}

\keywords{ultra fast, laser, pure spin current, pure valley current, valley, 2D material}

\maketitle

\section{Introduction}\label{sec1}

Spin and valley indices in two dimensional materials each endow the electron with a two-state freedom
~\cite{awschalom2002semiconductor,
meier2012optical}. This allows excitation of a novel type of current -- a {\it pure} current -- in which spin and valley momentum flow without charge transfer, minimizing the destructive Joule heating~\cite{hoffmann2007pure,huang2020pure} that inevitably accompanies charge currents. With the rise of two dimensional (2d) materials~\cite{gibertini_magnetic_2019}, and their potential for ultrafast light-driven spin/valley applications, there is now renewed impetus in exploring the generation and control of pure currents, with several proposed optical and non-optical approaches to their generation.

The latter include the spin Seebeck effect~\cite{uchida2008observation,uchida2008thermo,
ligeneration2014,yunonlinear2014,hugatetunable2023}, spin filters~\cite{danilov_optically_2018}, spin pumping~\cite{ding_direct_2020,hirsch1999spin,
governale2003pumping,mosendz2010quantifying}, and the spin Hall effect~\cite{pham2016ferromagnetic,
khang2018conductive}; all of these employ static thermal or electric potentials and thus possess spin current creation times ranging from several picoseconds to nanoseconds. Optical excitation via the photogalvanic and spin photogalvanic effects  achieves much faster generation times of several hundred femtoseconds to a picosecond, however this approach can require complex nano-device structures, involving for example edge states in nanoribbon or triangulene geometries, with inevitable sensitivity to structural disorder. Ultrafast laser pulses, on the other hand, offer direct coupling of the amplitude and phase of the electromagnetic waveform to the electron momenta and energy, permitting rich control over charge and current excitation on femtosecond times~\cite{higuchi2017light}. However despite apparent potential as a route to the ultrafast generation of pure currents, a general laser pulse protocol for their generation does not appear to exist.

In this work, we propose that a multi-pumped light pulse combining a single THz linearly polarized pulse with two gap tuned circularly polarized pulses of opposite helicity~\cite{shallcross2022ultrafast,sharma2023giant}, represents a general scheme for the ultrafast generation of 100\% pure spin and valley current. The linearly polarized component imparts crystal momenta $+\v k$ and $-\v k$ to the charge excitation created at the conjugate valleys by the circularly polarized components, resulting in a vanishing charge current but substantial spin (or valley) current. This approach we demonstrate offers (i) complete control over current direction and magnitude and (ii) adaptive robustness to imperfections in the lightwave forms, with highly irregular THz pulses -- as found in experiment -- capable of the generation of 100\% pure currents.
As the correlation between crystal momenta and spin/valley freedoms required of a pure current response is completely created by the light pulse, no special nano-device structure or band engineering is required. Our lightwave protocol is thus applicable to a wide range of valley active 2d materials, opening the way to a "lightwave pure current" physics at ultrafast femtosecond times.

\section{Results:}

\subsection{Pure spin current}

\begin{figure}[t!]
\centering\includegraphics[width=\textwidth]{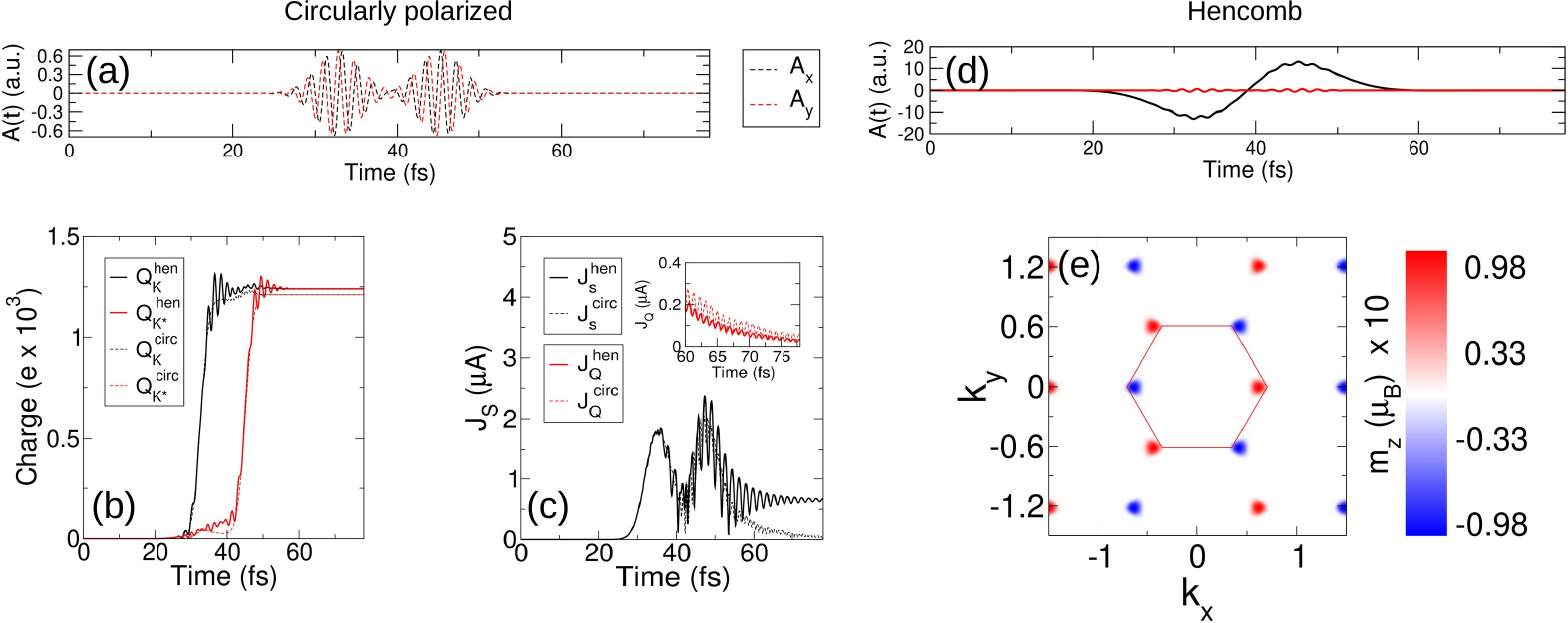}
\caption{{\it Light created pure spin current in a WSe$_2$ monolayer.}
A pulse train consisting of two temporally separated circularly polarized pulses of opposite helicity, panel (a), generates equal charge excitation at each valley, broken lines labelled $Q^{circ}$ in panel (c), with however zero spin $J_S$ or charge $J_Q$ residual current, broken lines panel (d) and the inset respectively. The addition of a monocycle of linearly polarized THz light, vector potential shown in panel (b), dramatically changes this situation. While the charge excitation at each valley is very similar to that created by the two bare circularly polarized pulses, panel (c) full lines, a substantial residual spin current is observed while the charge current remains vanishingly small, panel (d) full lines. The physics underlying this light creation of pure spin current is revealed by examination of the momentum and spin resolved excitation, panel (e), in which it is seen that the spin up and spin down excitations at conjugate valleys are displaced oppositely in momentum space by the light pulse.}
\label{fig1}
\end{figure}

We first consider the generation by laser light of a pure spin current in WSe$_2$; to this end we employ a minimal tight-binding model and evolution according to the von Neumann equation of motion (we use a decoherence time of 20~fs taken from a recent experiment\cite{heide_electronic_2021}). Full details of this methodology may be found in the Supplemental.

Generation of a pure spin current evidently requires excitation of both spin up and spin down charge, and hence charge excitation at both valleys of WSe$_2$. A sequence of gap tuned ($E_{gap} = 2.24$~eV) circularly polarized pulses of opposite helicity, Fig.~\ref{fig1}a, generates  successive charge excitation at each valley, broken lines Fig.~\ref{fig1}b, but zero overall spin or charge current, broken lines Fig.~\ref{fig1}c. This vanishing of the current follows as circularly polarized light excites a charge distribution inheriting the $C_3$ symmetry of the valley manifold, over which the microscopic current $\v j_{\v k}$ will necessarily integrate to zero.

To impart current to this charge excitation requires the introduction of second linearly polarized lightform. This pulse is time antisymmetric about the mid-point of the two circularly waveforms, see Fig.~\ref{fig1}d, with substantially sub-gap central frequency. Accordingly, this pulse excites only intra-band evolution of momenta according to the Bloch acceleration theorem $\v k(t) = \v k(0) - \v A(t)/c$ and, as the maxima and minima of the vector potential $\v A(t)$ coincide with the two oppositely polarized circularly polarized pulse components, this then generates an equal and opposite displacement in momentum space of the inter-band excitation created by these pulses.

Inspection of the spin momentum resolved excitation created by this pulse, Fig.~\ref{fig1}e, reveals the expected equal and opposite displacement of the K/K$^*$ valley spin up/down conduction band charge from the valley centre. This breaks the $C_3$ symmetry of the excitation, endowing each valley with equal and opposite current. Integrated over the Brillouin zone this pulse therefore yields a very similar charge excitation to the "bare" sequence of circularly polarized pulses, Fig.~\ref{fig1}b, but with dramatically different current response: the macroscopic charge current vanishes, while the macroscopic spin current remains finite, full lines in Fig.~\ref{fig1}c.

\begin{figure}[t!]
\centering\includegraphics[width=0.8\textwidth]{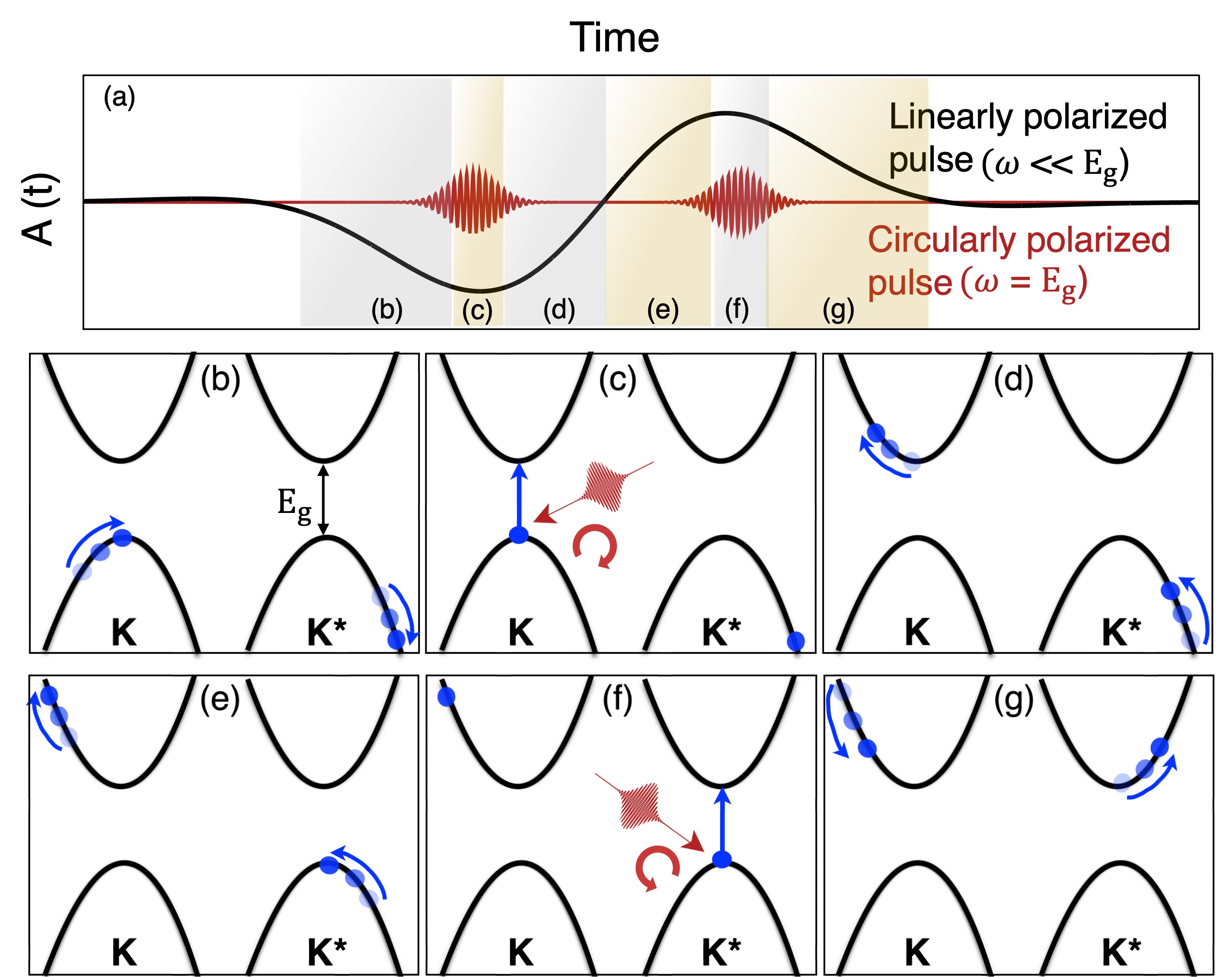}
\caption{Illustration of the excitation process underpinning the creating of pure spin current divided schematically into six steps, panels (b-g). Each of these panels corresponds to the shaded regions labelled in the vector potential plot shown in the topmost panel, panel (a). (a) Two initial opposite momenta at the K and K$^*$ valleys evolve, driven by the linearly polarized component of the pulse, intraband motion towards and away from the valley centres, respectively. (b) A circularly polarized pulse then generates excitation to the conduction band at the K valley, with no effect at the K$^*$ valley. (c) The second quarter cycle of the linearly polarized monocycle then returns the momenta to their initial values, resulting in a displaced charge excitation at K and no change at the K$^*$ valley. The second half cycle of the composite pulse repeats this process, but with changed sign of both the linearly polarized pulse vector potential and circularly polarized pulse helicity. This generates exactly the same excitation at the K$^*$ valley (with no change at K), but with opposite spin and momentum displacement, panels (d-g). Overall, the excitation process thus yields conjugate valleys with spin excitations displaced oppositely from the valley centres, and thus a vanishing charge current but a finite spin current.}
\label{fig2}
\end{figure}

The underlying mechanism of pure spin current generation employing this pulse design is illustrated schematically in Fig.~\ref{fig2}. The excitation process may be divided into six time steps, with each of these regions illustrated by a shaded region in the plot of the vector potential shown in the topmost panel (a). We consider two initial momenta, displaced at opposite sides of the two valley centres, panel (b). The linearly polarized pulse, sufficiently sub-gap in frequency not to generate interband excitation, evolves these initial momenta towards (K) and away (K$^*$) from the valley centres. Upon reaching the K valley centre the first ($\sigma^+$) circularly polarized pulse generates an interband transition, panel (c), before the second quarter cycle of linearly polarized light then returns the excited charge to its initial momenta, panel (d), but now in the conduction band. Thus at half-cycle the the K valley possesses excited charge displaced from the valley centre, while at the K$^*$ valley no excitation has occurred. The remaining half cycle of the pulse, panels (g-h), repeats this process but with the helicity of the circularly polarized pulse ($\sigma^-$) and the vector potential of the linearly polarized pulse now of opposite sign. The net result is thus to perform the exact same excitation at the K$^*$ valley, but with the momentum displacement of the excited charge exactly opposite to that of the K valley. Each valley thus possesses an equal and opposite current distribution consisting of spin up at the K valley and spin down at K$^*$, leading to zero charge but finite spin current, exactly as seen in the numerical results presented in Fig.~\ref{fig1}. 

\subsection{Direction and magnitude control}

\begin{figure}[t!]
\centering\includegraphics[width=0.9\textwidth]{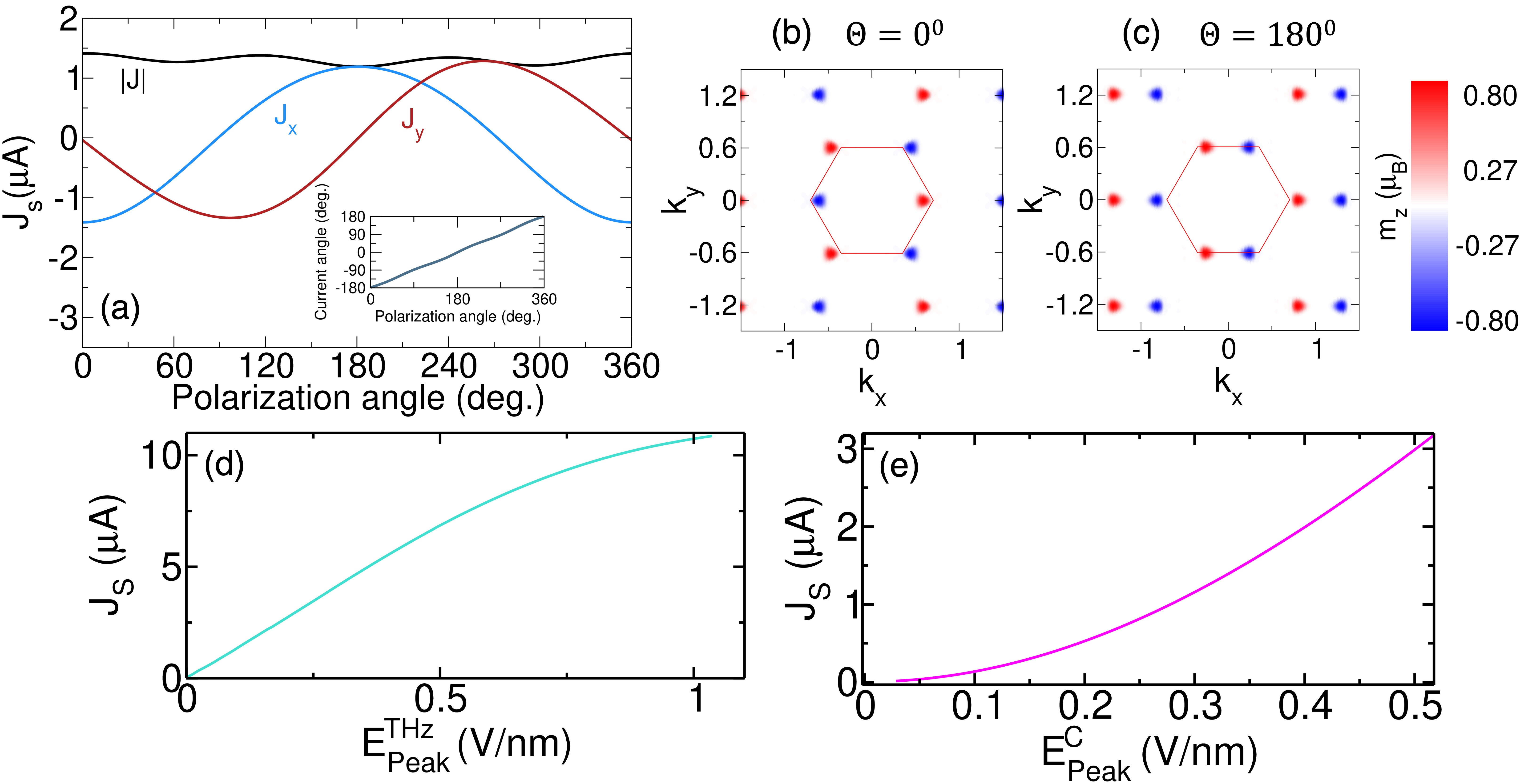}
\caption{\it{Complete control over the magnitude and direction of the spin current.} \rm{(a) Spin current shown as function of the polarization angle of linearly polarized THz pulse component; the angle of the spin current varies is equal to the polarization angle of the linearly polarized light. (b-c) Momentum resolved charge excitation, for two representative polarization angles of $\theta = 0^\circ$ and $\theta = 180^\circ$, respectively, with the color bar indicating the magnitude of the spin moment. (d) The spin current initially increases linearly with the peak electric field of linearly polarized light (E$_{\rm{Peak}}^{\rm{THz}}$) before saturating at higher values. (e) In contrast the spin current varies quadratically with the peak electric field of circularly polarized light pulse (E$_{\rm{Peak}}^{\rm{C}}$).}}
\label{fig3}
\end{figure}

Having established a mechanism for the generation of pure spin current, we now consider how current magnitude and direction can be controlled by light pulse parameters. Underlying the current distribution is the displacement of excited charge from the valley centres induced by the linearly polarized pulse component, and the current will therefore "inherit" magnitude and direction from the magnitude and direction of the polarization vector of this pulse, respectively. That this is so can be seen in Fig.~\ref{fig3}a, in which the current exhibits a near perfect sinusoidal variation with the polarization angle of the THz pulse, driven by a changing displacement of the valley spin distributions, illustrated for two representative cases of $\theta = 0^\circ$ and $\theta = 180^\circ$ in the momentum resolved excitations shown Fig.~\ref{fig3}b,c. (The small changes in the magnitude $|\v J_s|$ result from the valley $C_3$ symmetry~\cite{mccann2006landau,
cserti2007role,
kechedzhi2007influence,
nilsson2006electron}.) The magnitude of the linearly polarized component generates, Fig.~\ref{fig3}d, a linear increase in the pure spin current, an effect that has its origin in the linear increase Bloch velocity $\m\nabla_{\v k} \epsilon(\v k)$ away from the band edge. Saturation occurs when the $k^2$ behaviour near the band edge goes over to a near linear behaviour far from the valley centres. The pulse parameters of the sub-gap THz linearly polarized pulse component thus effect complete control over the current. The circularly polarized component, that generates interband excitation, can be tuned only to increase the current, Fig.~\ref{fig3}e, but cannot control current direction (the quadratic dependence indicating the dominance of one photon processes in the gap tuned interband excitation).

\subsection{Robustness to pulse shape}
     
\begin{figure}[t!]
\centering\includegraphics[width=0.9\textwidth]{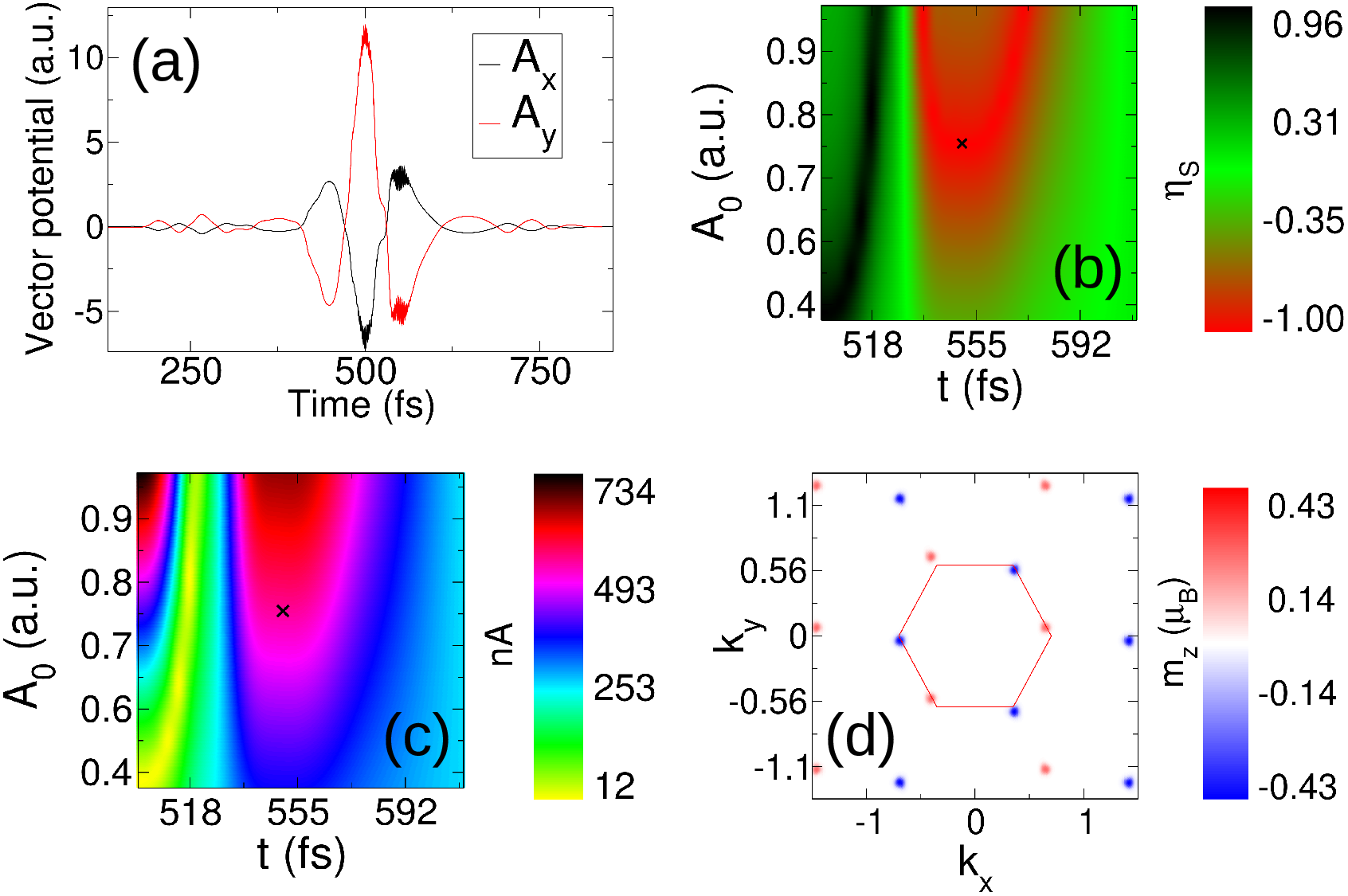}
\caption{{\it Robustness of multi-pump pure spin generation to THz envelope shape in WSe$_2$.} (a) A "noisy" linearly polarized THz waveform taken from experiment\cite{fulop_laser-driven_2020} is "dressed" by two circularly polarized pulses of opposite helicity, one coinciding with the central peak and a second with the shoulder peak. (b) By variation of the amplitude and time delay of this second pulse one can achieve a nearly 100\% pure spin current response which, as the delay time is reduced, evolves first into a 100\% spin polarized current and then to a pure charge current. (c) The spin current shows substantial variation, with a maxima near the "hot spot" region in which the charge current falls to zero. The momentum resolved excitation for the amplitude and delay time indicated by the crosses in panels (b,c); the different displacements from the high symmetry valley centres of the two spin distributions are compensated by the differing magnitude of the excitation at each centre, resulting in full compensation of the charge current and the emergence of a pure spin current.}
\label{fig4}
\end{figure}

The idealized THz monocycle thus far considered evidently represents a theoretical idealization, and to be experimentally useful this approach must remain valid even in the context of the "noisy" waveforms typical of few cycle pulses of infrared to THz light. To this end we take directly from experiment a pulse waveform for the linearly polarized component and combine this with two time delayed circularly polarized pulses of opposite helicity; the resulting vector potential is shown in Fig.~\ref{fig4}a. The time antisymmetry of the idealized monocycle is now absent, with the shoulder peaks (negative vector potential) considerably reduced in amplitude as compared to the central peak (positive vector potential). As we now show by variation of the amplitude and time delay of the second circularly polarized pulse a rich current response can be realized that includes a "hot spot" of pure spin currents.

To quantify this tuning procedure we may define a spin purity

\begin{equation}
    \eta_{S} = \frac{|J_Q| - |J_s|}{|J_Q| + |J_s|}
\end{equation}
where $J_Q$ corresponds to the total charge current i.e., $J_Q = J_{\uparrow}$ + $J_{\downarrow^*}$, where $J_\uparrow$/$J_\downarrow$ represent the spin up and spin down current respectively, with $J_s$ = $J_\uparrow$ - $J_\downarrow$ the spin current. Values of $\eta_S = -1,0,+1$ thus correspond respective to a pure spin current, a fully spin polarized charge current, and a pure charge current. In Fig.~\ref{fig4}b can be seen the variation of $\eta_S$ presented as a function of the time delay and amplitude $A_0$ of the second circularly polarized pulse. Variation of $A_0$ close to the shoulder peak (548~fs) yields a "hot region" of near 100\% pure spin current, with decrease in the time delay generating first a fully spin polarized charge current and then a pure charge current. These changes in the type of current response corresponds to the second circularly polarized pulse coinciding first with the node between the central and shoulder peak, and then moving into the central region itself. The momentum imbued to the second valley excitation in each of these cases is, respectively, zero (current at only only valley and thus a 100\% spin polarized current) and positive (same direction currents at each valley and thus a charge current response). 

The current generated for each of these delay time and amplitudes is presented in Fig.~\ref{fig4}c revealing a substantial spin current close to the pure spin hotspot. The cross in Fig.~\ref{fig4}b,c indicates the pulse in the pure spin current "hot spot" for which, in Fig.~\ref{fig4}d, we plot the momentum resolved excitation. As can be seen, the increased displacement from the K valley (red) as compared to the K$^\ast$ is compensated by the differing magnitudes of the two excitations, yielding overall the required charge current cancellation.

\subsection{Pure valley current in bilayer graphene} 
 
\begin{figure}[t!]
\centering\includegraphics[width=\textwidth]{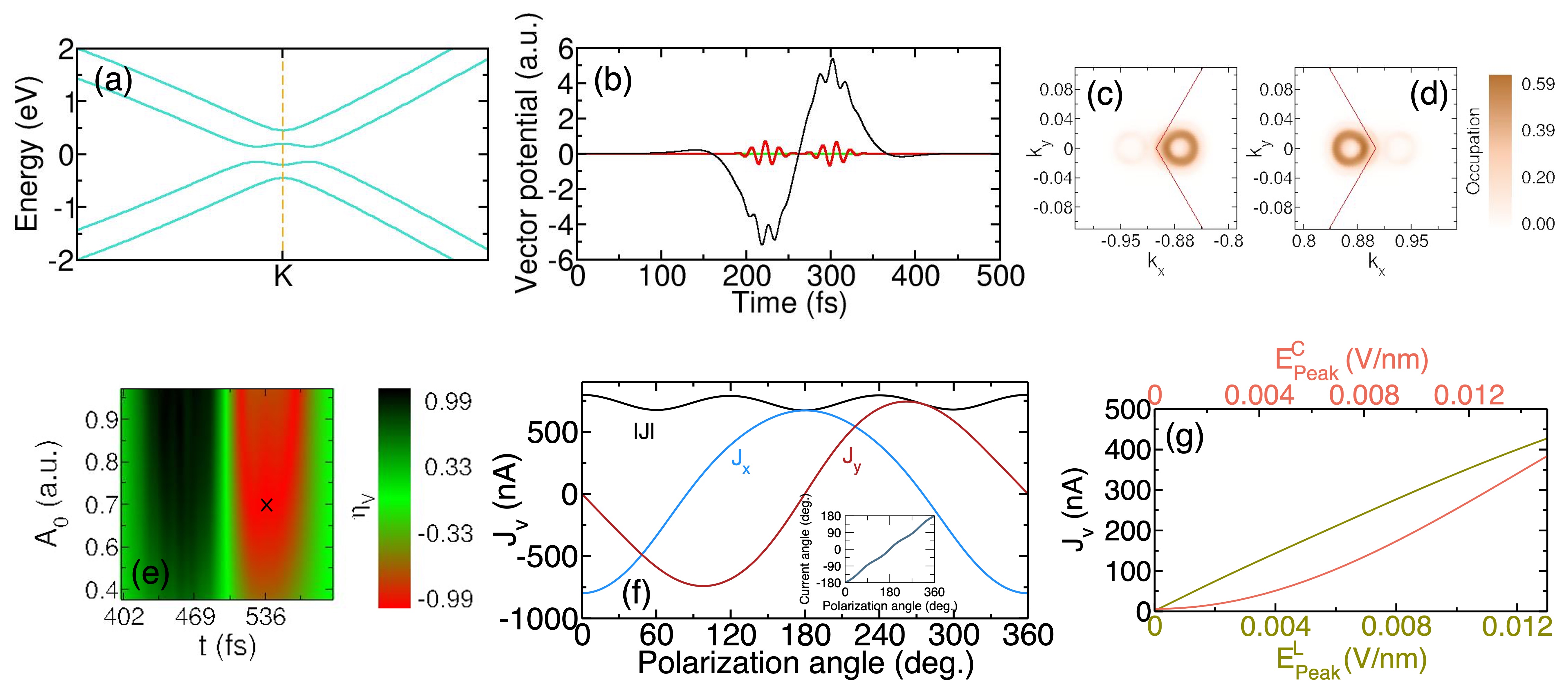}
\caption{\textit{Generation of pure valley current in bilayer graphene} (a) An applied interlayer symmetry breaking bias opens an infrared gap in bilayer graphene, allowing valley selective excitation of charge by circularly polarized light. A multi-pump pulse combining a monocycle of linearly polarized light with two opposite helicity circularly polarized waveforms, panel (b), generates identical charge excitations at each valley that, however, are displaced oppositely off the high symmetry valley centres, panels (c,d), yielding a finite valley current but zero charge current. Variation of the time delay and amplitude of the second circularly polarized pump pulse generates a rich current response, with both pure valley current ($\eta_v=-1$), 100\% valley polarized current ($\eta_v=0$), and pure charge current ($\eta_v=+1$) found, panel (e). The pure valley current "inherits" its direction and magnitude from the sub-gap linearly polarized pulse component, allowing full direction control via the polarization vector, panel (f), with magnitude control established via the amplitude of either the circularly or linearly polarized pulse component, panel (g).}
\label{fig5}
\end{figure}

Spin-orbit interaction generates inevitable dynamical loss of the spin character over time with valley currents, by contrast, suffering degradation of valley contrast only by large momentum intervalley scattering, that can, to some degree, be controlled by sample quality. Pure valley currents thus present an attractive alternative to pure spin currents. While the generation of valley polarized currents in graphene~\cite{yang2019chiral,
settnes2016graphene,luo2017optical}
and bilayer graphene~\cite{solomon2021valley} has received substantial attention, often employing deformation induced valley-contrasting gauge fields~\cite{gupta_straintronics_2019}, the generation of pure valley currents remains comparatively less explored~\cite{shimazaki_generation_2015}.

Here we show that the multi-pump pulse employed to generate pure spin current can also be employed to generate controllable pure valley currents in bilayer graphene. The low energy band structure of this material with a gap of 200~meV opened by the presence of an interlayer potential is presented in Fig.~\ref{fig5}a; details of our low energy Hamiltonian are provided in the supplemental. The valley degree of freedom can, as recently demonstrated in experiment, be addressed by infrared circularly polarized light, allowing the construction of an analogues pulse to that employed for creating pure spin current in WSe$_2$. This is shown in Fig.~\ref{fig5}b, along with the resulting momentum resolved charge excitation at each valley generated by this pulse, Fig.~\ref{fig5}c,d. The "ring" structure of the excitation results from the "Mexican hat" low energy manifold that can be seen in Fig.~\ref{fig5}a, with the equal and opposite displacement of these distributions at conjugate valleys clearly visible. 

Variation of the time delay and amplitude of the second circularly polarized pulse generates an analogous series of current types to those seen in WSe$_2$. Valley purity of the current can be defined by

\begin{equation}
    \eta_v = \frac{|J_Q| - |J_v|}{|J_Q| + |J_v|}
\end{equation}
where $J_Q = J_{K} + J_{K^*}$ is the total charge current and $J_v = J_{K} - J_{K^*}$ the valley current. Values of $\eta_v$ of -1, 0, and 1 correspond to a pure valley current, a fully valley polarized current, and a pure charge current respectively. As may be seen in Fig.~\ref{fig5}e, reducing the time delay of the second circularly polarized pulse send the current through a fully valley polarized state to a pure charge state, behaviour in exact analogy to that found in WSe$_2$. Finally we note that the complete control over pure current that our light protocol allows exists also for this pure valley current case. In Fig.~\ref{fig5}f we present the same direction control over the current with the angle of the polarization vector of the linearly polarized pulse component, while in Fig.~\ref{fig5}g is shown the linear and quadratic growth in the current magnitude with, respectively, the amplitude of the linearly and circularly polarized pulse components.
 
\section{Conclusion}

Strong field lasers couple light wave amplitude and phase to the fundamental electronic degrees of freedom of a solid, allowing rich control over macroscopic charge and spin excitations. The multi-pump pulse protocol described here "momentum dresses", via a THz envelope, the zero-current charge distribution created by circularly polarized light. This generates a very rich current response including both pure charge currents, fully polarized valley and spin currents, and pure valley and spin currents, with the particular current type determined by the time delay between the THz and gap-tuned pulse components. We have explored this in the context of the transition metal dichalcogenide WSe$_2$ and bilayer graphene, demonstrating respectively the generation of substantial pure spin and pure valley currents in these materials.

As the correlation between spin/valley freedoms and crystal momentum is created entirely by the light pulse, no special device design is called for. Furthermore, the internal time delay and amplitude freedoms of the pulse components imbues the multi-pump with significant robustness to irregularities -- inevitable in experiment -- that may occur in the THz envelope. Our approach thus offers a route towards 100\% pure spin and valley currents robust to both infelicities in pulse shape and requiring minimal material design, offering rich new possibilities for the ultrafast generation of pure spin and valley currents in 2d materials.

\backmatter

\bmhead{Supplementary information}
See Supplemental document for supporting content.

\bmhead{Acknowledgements}
Gill would like to thank DFG for funding through project-ID 328545488 TRR227 (project A04). Sharma would like to thank DFG for funding through project-ID 328545488 TRR227 (projects A04) and the Leibniz Professorin Program (SAW P118/2021). Shallcross would like to thank DFG for funding through project-ID 522036409 SH 498/7-1. The authors acknowledge the North-German Supercomputing Alliance (HLRN) for providing HPC resources that have contributed to the research results reported in this paper.

\section*{Declarations}

\begin{itemize}
\item Funding
DFG for funding through project-ID 328545488 TRR227 (project A04); DFG for funding through project-ID 328545488 TRR227 (projects A04); Leibniz Professorin Program (SAW P118/2021); DFG for funding through project-ID 522036409 SH 498/7-1. 
\item Conflict of interest/Competing interests :
The authors declare no conflicts of interest.
\item Ethics approval and consent to participate :
Not applicable.
\item Consent for publication :
All authors agree. 
\item Data availability :
Data underlying the results presented in this paper are not publicly available at this time but may be obtained from the authors upon reasonable request.
\item Materials availability :
Not applicable.
\item Code availability :
Available on reasonable request.
\item Author contribution :
S. Shallcross designed the project; S. Shallcross and D. Gill performed the tight-binding calculations; all authors contributed critically to the analysis and writing of the manuscript
\end{itemize}

\bibliography{deepika,current}

\begin{thebibliography}{10}
\expandafter\ifx\csname url\endcsname\relax
  \def\url#1{\burl{#1}}\fi
\expandafter\ifx\csname urlprefix\endcsname\relax\def\urlprefix{URL }\fi
\providecommand{\bibinfo}[2]{#2}
\providecommand{\eprint}[2][]{\url{#2}}
\providecommand{\doi}[1]{\url{https://doi.org/#1}}
\bibcommenthead

\bibitem{awschalom2002semiconductor}
\bibinfo{author}{Awschalom, D.} \& \bibinfo{author}{Loss, D.}
\newblock \emph{\bibinfo{title}{Semiconductor spintronics and quantum
  computation}}  (\bibinfo{publisher}{Springer Science \& Business Media},
  \bibinfo{year}{2002}).

\bibitem{meier2012optical}
\bibinfo{author}{Meier, F.} \& \bibinfo{author}{Zakharchenya, B.~P.}
\newblock \emph{\bibinfo{title}{Optical orientation}}
  (\bibinfo{publisher}{Elsevier}, \bibinfo{year}{2012}).

\bibitem{hoffmann2007pure}
\bibinfo{author}{Hoffmann, A.}
\newblock \bibinfo{title}{Pure spin-currents}.
\newblock \emph{\bibinfo{journal}{physica status solidi c}}
  \textbf{\bibinfo{volume}{4}}, \bibinfo{pages}{4236--4241}
  (\bibinfo{year}{2007}).

\bibitem{huang2020pure}
\bibinfo{author}{Huang, S.} \emph{et~al.}
\newblock \bibinfo{title}{Pure spin current phenomena}.
\newblock \emph{\bibinfo{journal}{Applied Physics Letters}}
  \textbf{\bibinfo{volume}{117}} (\bibinfo{year}{2020}).

\bibitem{gibertini_magnetic_2019}
\bibinfo{author}{Gibertini, M.}, \bibinfo{author}{Koperski, M.},
  \bibinfo{author}{Morpurgo, A.~F.} \& \bibinfo{author}{Novoselov, K.~S.}
\newblock \bibinfo{title}{Magnetic {2D} materials and heterostructures}.
\newblock \emph{\bibinfo{journal}{Nature Nanotechnology}}
  \textbf{\bibinfo{volume}{14}}, \bibinfo{pages}{408--419}
  (\bibinfo{year}{2019}).
\newblock \urlprefix\url{https://www.nature.com/articles/s41565-019-0438-6}.
\newblock \bibinfo{note}{Number: 5 Publisher: Nature Publishing Group}.

\bibitem{uchida2008observation}
\bibinfo{author}{Uchida, K.-I.} \emph{et~al.}
\newblock \bibinfo{title}{Observation of the spin seebeck effect}.
\newblock \emph{\bibinfo{journal}{Nature}} \textbf{\bibinfo{volume}{455}},
  \bibinfo{pages}{778--781} (\bibinfo{year}{2008}).

\bibitem{uchida2008thermo}
\bibinfo{author}{Uchida, K.} \emph{et~al.}
\newblock \bibinfo{title}{Thermo-spintronics: A new direction for spins}.
\newblock \emph{\bibinfo{journal}{Nature}} \textbf{\bibinfo{volume}{455}},
  \bibinfo{pages}{778} (\bibinfo{year}{2008}).

\bibitem{ligeneration2014}
\bibinfo{author}{Li, P.} \emph{et~al.}
\newblock \bibinfo{title}{Generation of pure spin currents via spin seebeck
  effect in self-biased hexagonal ferrite thin films}.
\newblock \emph{\bibinfo{journal}{Applied Physics Letters}}
  \textbf{\bibinfo{volume}{105}}, \bibinfo{pages}{242412}.
\newblock \urlprefix\url{https://doi.org/10.1063/1.4904479}.

\bibitem{yunonlinear2014}
\bibinfo{author}{Yu, H.}, \bibinfo{author}{Wu, Y.}, \bibinfo{author}{Liu,
  G.-B.}, \bibinfo{author}{Xu, X.} \& \bibinfo{author}{Yao, W.}
\newblock \bibinfo{title}{Nonlinear valley and spin currents from fermi pocket
  anisotropy in 2d crystals}.
\newblock \emph{\bibinfo{journal}{Phys. Rev. Lett.}}
  \textbf{\bibinfo{volume}{113}}, \bibinfo{pages}{156603}.
\newblock
  \urlprefix\url{https://link.aps.org/doi/10.1103/PhysRevLett.113.156603}.
\newblock \bibinfo{note}{Publisher: American Physical Society}.

\bibitem{hugatetunable2023}
\bibinfo{author}{Hu, Y.}, \bibinfo{author}{Liu, S.}, \bibinfo{author}{Huang,
  J.}, \bibinfo{author}{Li, X.} \& \bibinfo{author}{Li, Q.}
\newblock \bibinfo{title}{Gate-tunable spin seebeck effect and pure spin
  current generation in molecular junctions based on bipolar magnetic
  molecules}.
\newblock \emph{\bibinfo{journal}{Nano Lett.}} \textbf{\bibinfo{volume}{23}},
  \bibinfo{pages}{7890--7896}.
\newblock \urlprefix\url{https://doi.org/10.1021/acs.nanolett.3c01702}.
\newblock \bibinfo{note}{Publisher: American Chemical Society}.

\bibitem{danilov_optically_2018}
\bibinfo{author}{Danilov, A.~P.} \emph{et~al.}
\newblock \bibinfo{title}{Optically excited spin pumping mediating collective
  magnetization dynamics in a spin valve structure}.
\newblock \emph{\bibinfo{journal}{Phys. Rev. B}} \textbf{\bibinfo{volume}{98}},
  \bibinfo{pages}{060406}.
\newblock \urlprefix\url{https://link.aps.org/doi/10.1103/PhysRevB.98.060406}.
\newblock \bibinfo{note}{Publisher: American Physical Society}.

\bibitem{ding_direct_2020}
\bibinfo{author}{Ding, J.} \emph{et~al.}
\newblock \bibinfo{title}{Direct observation of spin accumulation in cu induced
  by spin pumping}.
\newblock \emph{\bibinfo{journal}{Phys. Rev. Res.}}
  \textbf{\bibinfo{volume}{2}}, \bibinfo{pages}{013262}.
\newblock
  \urlprefix\url{https://link.aps.org/doi/10.1103/PhysRevResearch.2.013262}.
\newblock \bibinfo{note}{Publisher: American Physical Society}.

\bibitem{hirsch1999spin}
\bibinfo{author}{Hirsch, J.}
\newblock \bibinfo{title}{Spin hall effect}.
\newblock \emph{\bibinfo{journal}{Physical review letters}}
  \textbf{\bibinfo{volume}{83}}, \bibinfo{pages}{1834} (\bibinfo{year}{1999}).

\bibitem{governale2003pumping}
\bibinfo{author}{Governale, M.}, \bibinfo{author}{Taddei, F.} \&
  \bibinfo{author}{Fazio, R.}
\newblock \bibinfo{title}{Pumping spin with electrical fields}.
\newblock \emph{\bibinfo{journal}{Physical Review B}}
  \textbf{\bibinfo{volume}{68}}, \bibinfo{pages}{155324}
  (\bibinfo{year}{2003}).

\bibitem{mosendz2010quantifying}
\bibinfo{author}{Mosendz, O.} \emph{et~al.}
\newblock \bibinfo{title}{Quantifying spin hall angles from spin pumping:
  Experiments and theory}.
\newblock \emph{\bibinfo{journal}{Physical review letters}}
  \textbf{\bibinfo{volume}{104}}, \bibinfo{pages}{046601}
  (\bibinfo{year}{2010}).

\bibitem{pham2016ferromagnetic}
\bibinfo{author}{Pham, V.~T.} \emph{et~al.}
\newblock \bibinfo{title}{Ferromagnetic/nonmagnetic nanostructures for the
  electrical measurement of the spin hall effect}.
\newblock \emph{\bibinfo{journal}{Nano letters}} \textbf{\bibinfo{volume}{16}},
  \bibinfo{pages}{6755--6760} (\bibinfo{year}{2016}).

\bibitem{khang2018conductive}
\bibinfo{author}{Khang, N. H.~D.}, \bibinfo{author}{Ueda, Y.} \&
  \bibinfo{author}{Hai, P.~N.}
\newblock \bibinfo{title}{A conductive topological insulator with large spin
  hall effect for ultralow power spin--orbit torque switching}.
\newblock \emph{\bibinfo{journal}{Nature materials}}
  \textbf{\bibinfo{volume}{17}}, \bibinfo{pages}{808--813}
  (\bibinfo{year}{2018}).

\bibitem{higuchi2017light}
\bibinfo{author}{Higuchi, T.}, \bibinfo{author}{Heide, C.},
  \bibinfo{author}{Ullmann, K.}, \bibinfo{author}{Weber, H.~B.} \&
  \bibinfo{author}{Hommelhoff, P.}
\newblock \bibinfo{title}{Light-field-driven currents in graphene}.
\newblock \emph{\bibinfo{journal}{Nature}} \textbf{\bibinfo{volume}{550}},
  \bibinfo{pages}{224--228} (\bibinfo{year}{2017}).

\bibitem{shallcross2022ultrafast}
\bibinfo{author}{Shallcross, S.}, \bibinfo{author}{Li, Q.},
  \bibinfo{author}{Dewhurst, J.}, \bibinfo{author}{Sharma, S.} \&
  \bibinfo{author}{Elliott, P.}
\newblock \bibinfo{title}{Ultrafast optical control over spin and momentum in
  solids}.
\newblock \emph{\bibinfo{journal}{Applied Physics Letters}}
  \textbf{\bibinfo{volume}{120}} (\bibinfo{year}{2022}).

\bibitem{sharma2023giant}
\bibinfo{author}{Sharma, S.}, \bibinfo{author}{Gill, D.} \&
  \bibinfo{author}{Shallcross, S.}
\newblock \bibinfo{title}{Giant and controllable valley currents in graphene by
  double pumped thz light}.
\newblock \emph{\bibinfo{journal}{Nano Letters}} \textbf{\bibinfo{volume}{23}},
  \bibinfo{pages}{10305--10310} (\bibinfo{year}{2023}).

\bibitem{heide_electronic_2021}
\bibinfo{author}{Heide, C.} \emph{et~al.}
\newblock \bibinfo{title}{Electronic {Coherence} and {Coherent} {Dephasing} in
  the {Optical} {Control} of {Electrons} in {Graphene}}.
\newblock \emph{\bibinfo{journal}{Nano Letters}} \textbf{\bibinfo{volume}{21}},
  \bibinfo{pages}{9403--9409} (\bibinfo{year}{2021}).
\newblock \bibinfo{note}{Publisher: American Chemical Society}.

\bibitem{mccann2006landau}
\bibinfo{author}{McCann, E.} \& \bibinfo{author}{Fal’ko, V.~I.}
\newblock \bibinfo{title}{Landau-level degeneracy and quantum hall effect in a
  graphite bilayer}.
\newblock \emph{\bibinfo{journal}{Physical review letters}}
  \textbf{\bibinfo{volume}{96}}, \bibinfo{pages}{086805}
  (\bibinfo{year}{2006}).

\bibitem{cserti2007role}
\bibinfo{author}{Cserti, J.}, \bibinfo{author}{Csord{\'a}s, A.} \&
  \bibinfo{author}{D{\'a}vid, G.}
\newblock \bibinfo{title}{Role of the trigonal warping on the minimal
  conductivity of bilayer graphene}.
\newblock \emph{\bibinfo{journal}{Physical review letters}}
  \textbf{\bibinfo{volume}{99}}, \bibinfo{pages}{066802}
  (\bibinfo{year}{2007}).

\bibitem{kechedzhi2007influence}
\bibinfo{author}{Kechedzhi, K.}, \bibinfo{author}{Fal’ko, V.~I.},
  \bibinfo{author}{McCann, E.} \& \bibinfo{author}{Altshuler, B.}
\newblock \bibinfo{title}{Influence of trigonal warping on interference effects
  in bilayer graphene}.
\newblock \emph{\bibinfo{journal}{Physical review letters}}
  \textbf{\bibinfo{volume}{98}}, \bibinfo{pages}{176806}
  (\bibinfo{year}{2007}).

\bibitem{nilsson2006electron}
\bibinfo{author}{Nilsson, J.}, \bibinfo{author}{Neto, A.~C.},
  \bibinfo{author}{Peres, N.} \& \bibinfo{author}{Guinea, F.}
\newblock \bibinfo{title}{Electron-electron interactions and the phase diagram
  of a graphene bilayer}.
\newblock \emph{\bibinfo{journal}{Physical Review B}}
  \textbf{\bibinfo{volume}{73}}, \bibinfo{pages}{214418}
  (\bibinfo{year}{2006}).

\bibitem{fulop_laser-driven_2020}
\bibinfo{author}{Fülöp, J.~A.}, \bibinfo{author}{Tzortzakis, S.} \&
  \bibinfo{author}{Kampfrath, T.}
\newblock \bibinfo{title}{Laser-{Driven} {Strong}-{Field} {Terahertz}
  {Sources}}.
\newblock \emph{\bibinfo{journal}{Advanced Optical Materials}}
  \textbf{\bibinfo{volume}{8}}, \bibinfo{pages}{1900681}
  (\bibinfo{year}{2020}).
\newblock
  \urlprefix\url{https://onlinelibrary.wiley.com/doi/abs/10.1002/adom.201900681}.
\newblock \bibinfo{note}{\_eprint:
  https://onlinelibrary.wiley.com/doi/pdf/10.1002/adom.201900681}.

\bibitem{yang2019chiral}
\bibinfo{author}{Yang, Z.}, \bibinfo{author}{Aghaeimeibodi, S.} \&
  \bibinfo{author}{Waks, E.}
\newblock \bibinfo{title}{Chiral light-matter interactions using spin-valley
  states in transition metal dichalcogenides}.
\newblock \emph{\bibinfo{journal}{Optics express}}
  \textbf{\bibinfo{volume}{27}}, \bibinfo{pages}{21367--21379}
  (\bibinfo{year}{2019}).

\bibitem{settnes2016graphene}
\bibinfo{author}{Settnes, M.}, \bibinfo{author}{Power, S.~R.},
  \bibinfo{author}{Brandbyge, M.} \& \bibinfo{author}{Jauho, A.-P.}
\newblock \bibinfo{title}{Graphene nanobubbles as valley filters and beam
  splitters}.
\newblock \emph{\bibinfo{journal}{Physical review letters}}
  \textbf{\bibinfo{volume}{117}}, \bibinfo{pages}{276801}
  (\bibinfo{year}{2016}).

\bibitem{luo2017optical}
\bibinfo{author}{Luo, M.} \& \bibinfo{author}{Li, Z.}
\newblock \bibinfo{title}{Optical excitation of valley and spin currents of
  chiral edge states in graphene with rashba spin-orbital coupling}.
\newblock \emph{\bibinfo{journal}{Physical Review B}}
  \textbf{\bibinfo{volume}{96}}, \bibinfo{pages}{165424}
  (\bibinfo{year}{2017}).

\bibitem{solomon2021valley}
\bibinfo{author}{Solomon, F.} \& \bibinfo{author}{Power, S.~R.}
\newblock \bibinfo{title}{Valley current generation using biased bilayer
  graphene dots}.
\newblock \emph{\bibinfo{journal}{Physical Review B}}
  \textbf{\bibinfo{volume}{103}}, \bibinfo{pages}{235435}
  (\bibinfo{year}{2021}).

\bibitem{gupta_straintronics_2019}
\bibinfo{author}{Gupta, R.}, \bibinfo{author}{Rost, F.},
  \bibinfo{author}{Fleischmann, M.}, \bibinfo{author}{Sharma, S.} \&
  \bibinfo{author}{Shallcross, S.}
\newblock \bibinfo{title}{Straintronics beyond homogeneous deformation}.
\newblock \emph{\bibinfo{journal}{Physical Review B}}
  \textbf{\bibinfo{volume}{99}}, \bibinfo{pages}{125407}
  (\bibinfo{year}{2019}).
\newblock \urlprefix\url{https://link.aps.org/doi/10.1103/PhysRevB.99.125407}.

\bibitem{shimazaki_generation_2015}
\bibinfo{author}{Shimazaki, Y.} \emph{et~al.}
\newblock \bibinfo{title}{Generation and detection of pure valley current by
  electrically induced {Berry} curvature in bilayer graphene}.
\newblock \emph{\bibinfo{journal}{Nature Physics}}
  \textbf{\bibinfo{volume}{11}}, \bibinfo{pages}{1032--1036}
  (\bibinfo{year}{2015}).
\newblock \urlprefix\url{https://www.nature.com/articles/nphys3551}.
\newblock \bibinfo{note}{Number: 12 Publisher: Nature Publishing Group}.

\end{thebibliography}

\end{document}